# Machine learning for the design and prediction of soft-magnetic electromagnetic shielding FeCo-based alloys in laser cladding

Luting Wang[a,c], Suiyuan Chen[a*], Xiancheng Zhu[a], Zhiqing Fang[a], Mei Wang[b], Yifan Li[c], Lei Shen[c]

[a] Key Laboratory for Anisotropy and Texture of Materials, Ministry of Education, School of Materials Science and Engineering, Northeastern University, Shenyang 110819, Liaoning, China. *Email: chensy@atm.neu.edu.cn.

[b] Shenyang Dalu Laser Technology Co. Ltd., China.

[c] The Department of Mechanical Engineering, National University of Singapore, Singapore, 117575

**Abstract**

Electromagnetic shielding materials play a pivotal role in both aerospace applications and daily life. However, their design and manufacturing still face persistent challenges. Machine learning demonstrates significant potential in accelerating material development and compositions optimization. Furthermore, laser additive manufacturing provides powerful technical support for fabricating multi-component, multifunctional electromagnetic shielding materials with tailored properties. In this study, the multiple machine learning strategies have been proposed, based on experimental derivation and soft magnetic material databases, to accelerate the design of multifunctional FeCo-based alloys for electromagnetic interference (EMI) shielding within an almost infinite compositional space. This work presents a novel approach for the rapid and automated discovery of multifunctional alloys with optimized EMI shielding effectiveness, as well as enhanced magnetic and electrical properties.

## 1. Introduction

Laser additive manufacturing, known for its compatibility with a wide range of materials, has garnered significant attention due to its high productivity and capabilities for in-situ fabrication and repair [1-3]. The EMI materials design and the precise control of laser parameters are critical to ensuring the quality and performance of the manufactured components [4, 5]. Currently, little special mathematical model exists to directly correlate initial experimental parameters with the final properties. As a result, shortening production cycles and enabling accurate prediction of EMI performance remain pressing challenges in the advancement of industrial-scale laser additive manufacturing.

To overcome these obstacles, with the development of electromagnetic interference shielding alloys, we have reached a stage where multiple elements combined laser parameters are used in 3D printing. In this study, we focused on the design of multi-functional soft magnetic alloys with EMI shielding efficiency for several reasons: (i) a high demand exists for multi-functional soft magnetic alloys to serve the electromagnetic protection of military communications, radar detection and electronic devices [6-8]; (ii) alternative elements and regulable microstructures (e.g., soft magnetic phase and hard magnetic phase with various shape and size) [9, 10]; (iii) each element has an important influence on the shielding performance, but the cooperative relationship of multiple elements is not clear at present; (iv) laser cladding is a critical part in 3D printing, manufacturing parameters significantly affect both the printability and functional properties of the alloys [11-13]. However, comprehensive datasets on processing parameters for multifunctional EMI shielding alloys are currently lacking; (v) The pursuit of high efficiency in EMI shielding requires a new generation of multi-functional materials designed through integrated, data-driven approaches.

Due to limitations in dataset, the discovery of alloys with promising EMI shielding properties remains a significant challenge. To address this, we propose a novel approach for accelerating the design and discovery of EMI shielding alloys. This strategy integrates key characteristics of soft magnetic alloys with machine learning (ML) techniques [14, 15] and Φ-SO (a Physical Symbolic Optimization framework that leverages deep reinforcement learning to recover analytical symbolic expressions from physical data by learning unit constraints) [16]. Following the construction of experimental EMI materials database, Φ-SO is employed to train artificial neural networks capable of generating accurate, interpretable expression. This not only improves prediction accuracy but also enhances the transparency and explainability of the machine learning process.

The machine learning and Φ-SO methods involve several key steps (as shown in Fig. 1): data collection, physics-informed screening, feature extraction and model development, and expression generation. Due to plentiful combinations of compositions, microstructures, and processing parameters for EMI shielding materials, a major challenge lies in directly designing these variables while simultaneously tailoring functional properties through predictive modeling. To address this, we employ multiple ML models, including Decision Tree, Random Forest, Support Vector Regression (SVR), Ridge Regression, K-Nearest Neighbors (K-NN), and Gradient Boosting, to guide the design of EMI shielding alloys with soft magnetic properties.

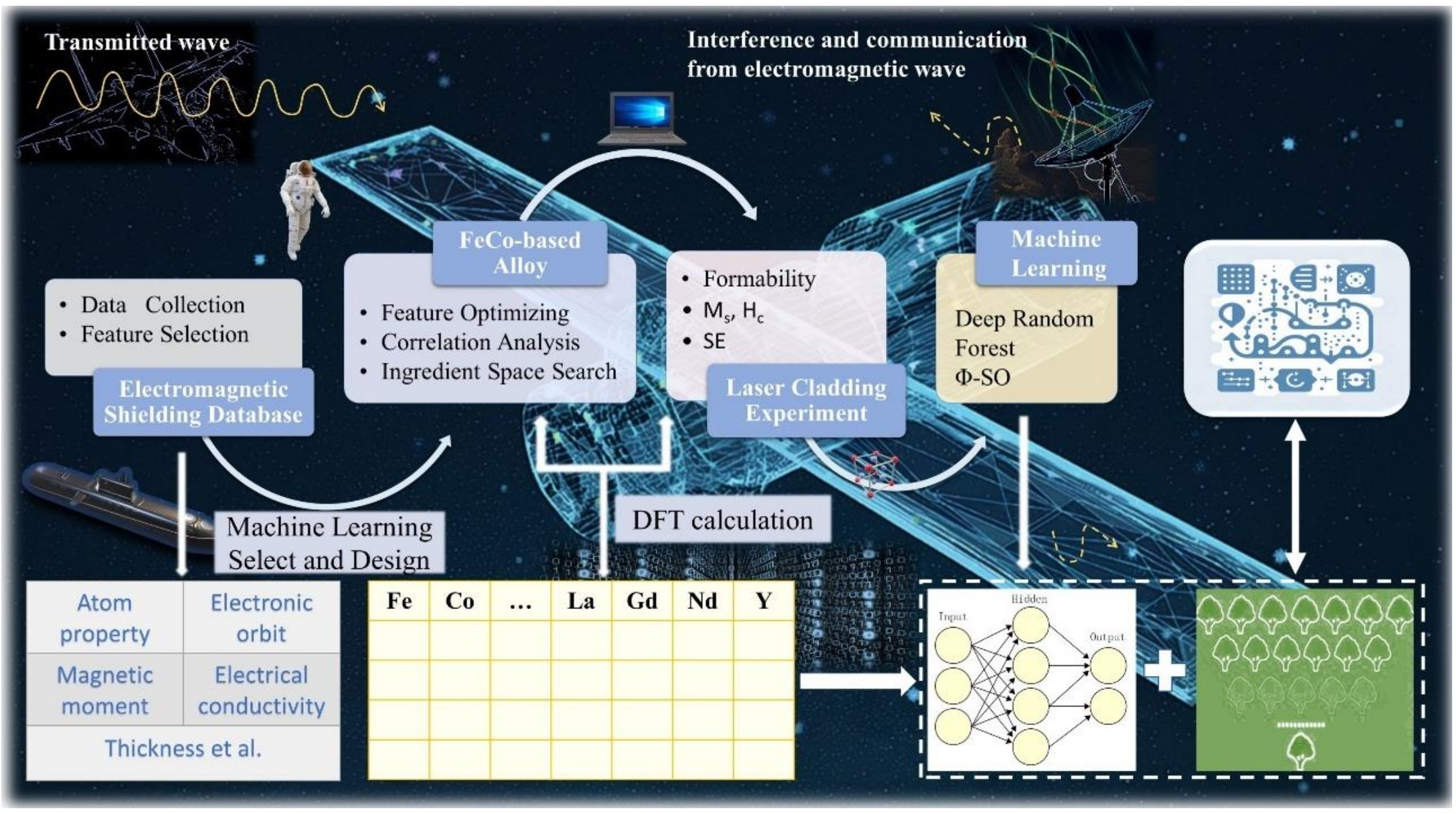

Fig. 1. Diagram of machine learning and experimental design framework

## 2. Data collection and collation of electromagnetic shielding materials

### 2.1. Feature selection and optimization

There are common evaluation standards for electromagnetic shielding performance across various materials [17, 18]. By collecting relevant characteristics of different electromagnetic shielding materials, not only can a comprehensive set of features for evaluating electromagnetic shielding performance be derived, but important characteristics related to alloy-type electromagnetic shielding materials can also be extracted [19-27]. Based on the collected and organized data of electromagnetic shielding materials, it is found that the vast majority of high-performance electromagnetic shielding materials contain different types and proportions of metallic elements, indicating that metallic elements are an important component of electromagnetic shielding materials and have a significant impact on shielding performance [28-31]. However, some elements may have adverse effects on the environment or health, such as mercury, cadmium, lead, and arsenic, therefore, these elements are not within the scope of selection for this study.

Through collecting relevant properties from a wide range of electromagnetic shielding materials, it is possible to derive a comprehensive set of features for performance evaluation, while also identifying key characteristics specific to alloy-based shielding materials [32].

Analysis of the compiled dataset reveals that most high-performance electromagnetic shielding materials incorporate various types and proportions of metallic elements. However, certain elements, such as mercury, cadmium, lead, and arsenic—pose environmental and health risks [33, 34]. As a result, these hazardous elements are excluded from consideration in this study. Table 1 summarizes the types of elements, their content ranges, and theoretical conductivities derived from the comprehensive database of electromagnetic shielding materials.

Table 1 The proportion and electrical conductivity of elements in the dataset [17-32]

| Element | Maximum (wt. %) | Minimum (wt. %) | Conductivity (S/m) |
|---|---|---|---|
| Ag | 6.0 | 0 | $6.3\times 10^{7}$ |
| Cu | 100.0 | 0 | $6.0\times 10^{7}$ |
| Al | 100.0 | 0 | $3.5\times 10^{7}$ |
| Mg | 95.0 | 0 | $2.3\times 10^{7}$ |
| Zn | 50.0 | 0 | $1.7\times 10^{7}$ |
| Ni | 1.1 | 0 | $1.4\times 10^{7}$ |
| Li | 8.9 | 0 | $1.1\times 10^{7}$ |
| Fe | 90.0 | 0 | $1.0\times 10^{7}$ |
| Pt | 20.0 | 0 | $9.4\times 10^{6}$ |
| Mo | 10.0 | 0 | $1.9\times 10^{7}$ |
| Ti/Sn | 5.0 | 0 | $9.2\times 10^{6}$ |
| Stainless steel | 100.0 | 0 | $1.5\times 10^{6}$ |
| Co | 30.0 | 0 | $1.7\times 10^{5}$ |
| Zr | 10.3 | 0 | $4.2\times 10^{6}$ |
| Si | 90.0 | 0 | $1.6\times 10^{-3}$ |
| SiC | 100.0 | 0 | $1.0\times 10^{-12}$ |
| B | 10.0 | 0 | $1.0\times 10^{-12}$ |
| Glass | 100.0 | 0 | $1.0\times 10^{-13}$ |
| Polyethylene terephthalate | 100.0 | 0 | $1.0\times 10^{-14}$ |
| Wood (dry) | 11.0 | 0 | $1.0\times 10^{-15}$ |
| Polylactic acid | 100.0 | 0 | $1.0\times 10^{-16}$ |
| Graphene | 25.0 | 0 | $1.0\times 10^{6}$ |
| Carbon nano tube | 50.0 | 0 | $1.0\times 10^{6}$ |

Electrical conductivity is positively correlated with electromagnetic reflection loss efficiency; higher reflection losses contribute to improved overall shielding effectiveness [35-37]. By analyzing conductivity data in conjunction with the practical requirements of laser cladding experiments and referencing relevant electromagnetic shielding phases extracted from the materials database (as illustrated in Fig. 2) [17-32], the primary elemental compositions for laser cladding experiments were selected from Table 1. The compositional ranges of these selected elements were subsequently refined and optimized to ensure desirable performance and printability.

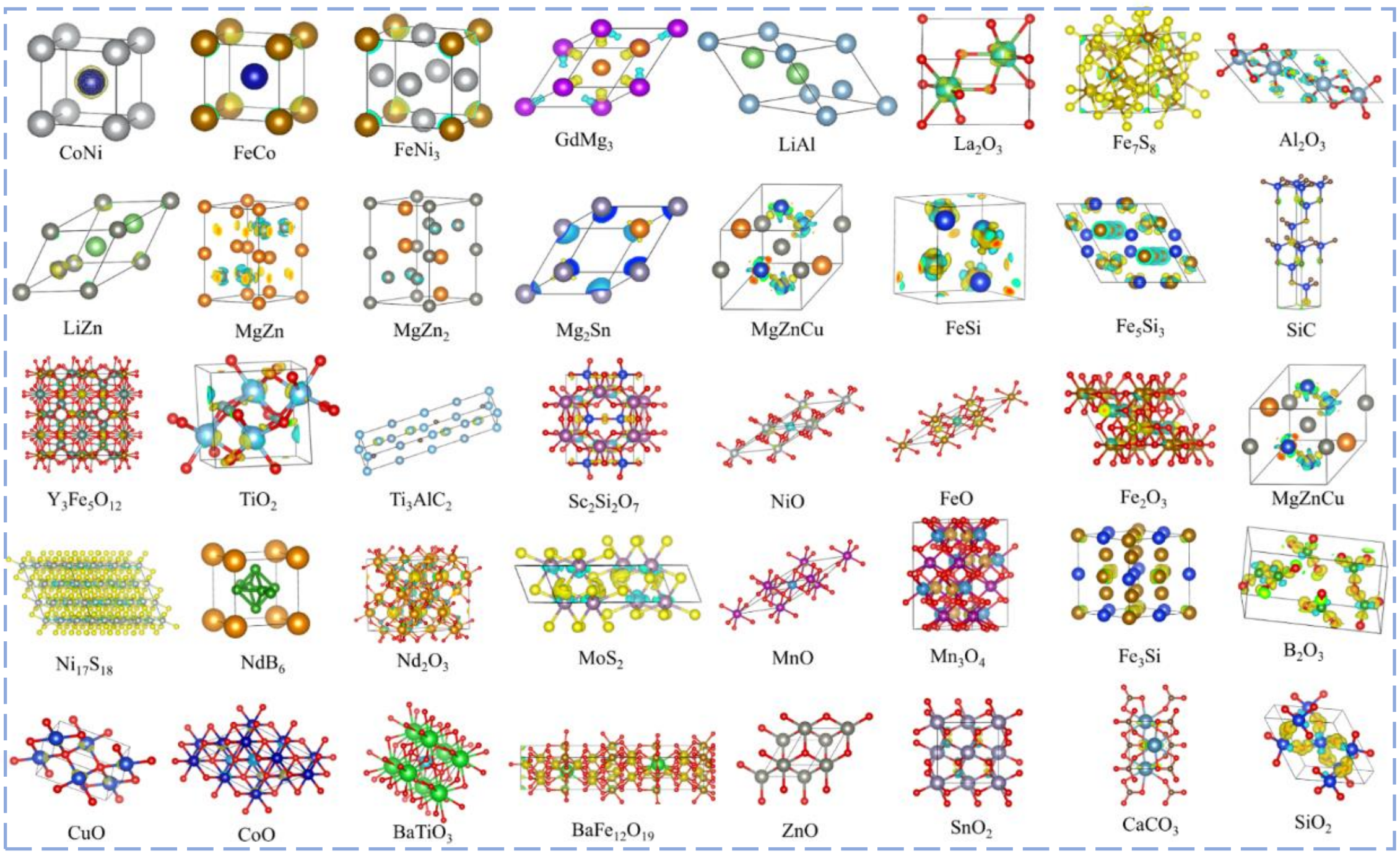

Fig. 2. Phases in electromagnetic shielding interference materials by the first principle calculation

Based on the collected data on electromagnetic shielding materials, the following key features were extracted: P (atomic percentage of the added material, at. %), T (sample thickness, mm), S (structural characteristics of the sample), M (types of elements), F (structural form of the added elements: powder, hollow tubular, solid wire, pure solid, or polymorphic mixture), S2 (minimum structural size of the added elements), C (measured conductivity of the sample), Cf (theoretical conductivity of the added elements), Cb (theoretical conductivity of the matrix), and SE (shielding effectiveness). Due to the large numerical disparity between conductivity-related features and other variables, data preprocessing was required prior to analysis. The processed results are presented in Fig. 3.

Fig. 3(a) presents the importance ranking of all features influencing electromagnetic shielding performance, while Fig. 3(b) illustrates the spatial distribution of several key features, with the size of the spheres representing the values of the matrix's theoretical conductivity (Cb). The analysis reveals that the theoretical conductivity of the unprocessed matrix (Cb) and the actual conductivity of the processed material (C) are critical factors affecting shielding performance. Additionally, sample thickness (T) also plays a significant role. These findings are consistent with established factors influencing electromagnetic shielding effectiveness, supporting the reliability of the collected material features, the rationality of the selected descriptors, and the accuracy of the machine learning model employed for feature importance analysis.

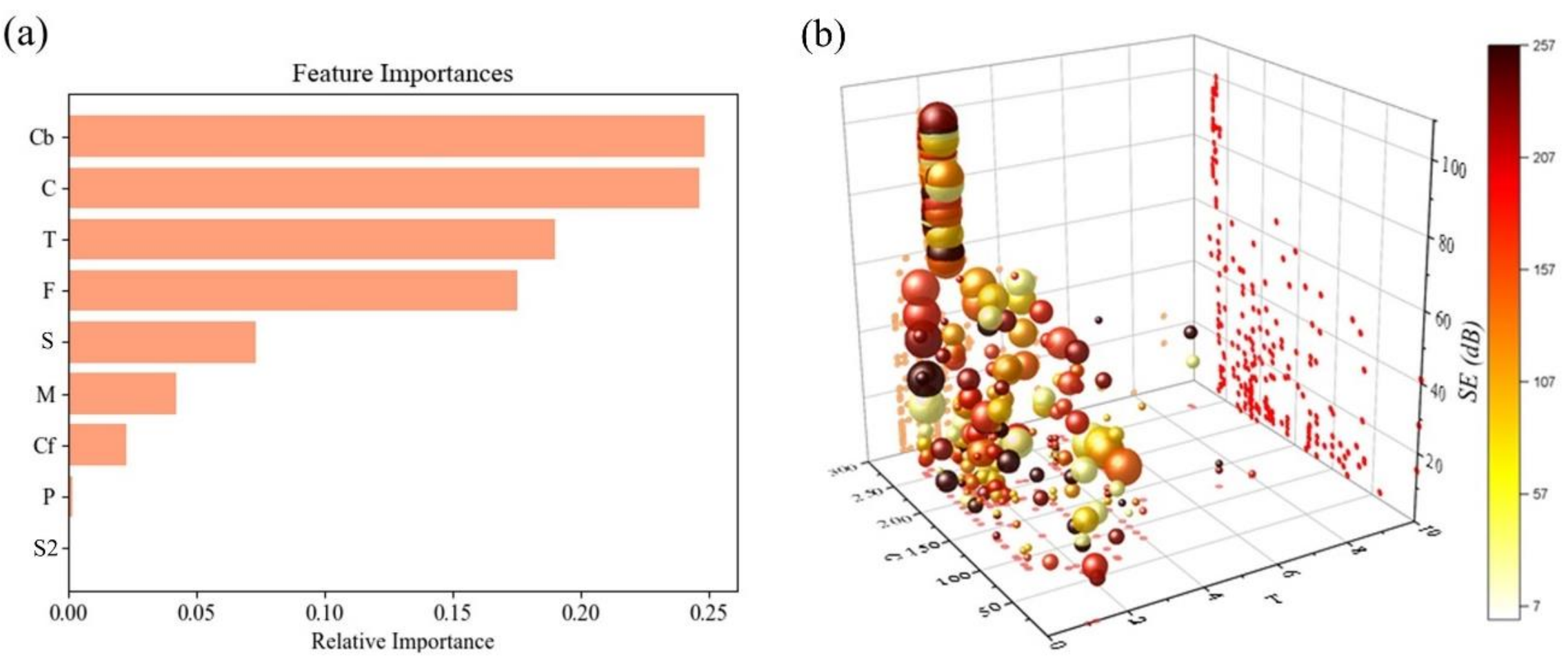

Fig. 3. Feature importance and major features distribution of EMI shielding materials

(a) Ranking of feature importance; (b) Spatial distribution of key features

Based on the feature importance ranking results, the feature S2 (minimum structural size of the added elements) was found to have relatively low relevance to shielding performance. As a result, S2 was excluded during the feature selection process, and the remaining features were retained for subsequent correlation analysis of electromagnetic shielding performance. Fig. 4 presents a comprehensive correlation analysis, including Pearson correlation coefficients, Spearman rank correlation coefficients, Kendall Tau rank correlation coefficients, and distance correlation matrices.

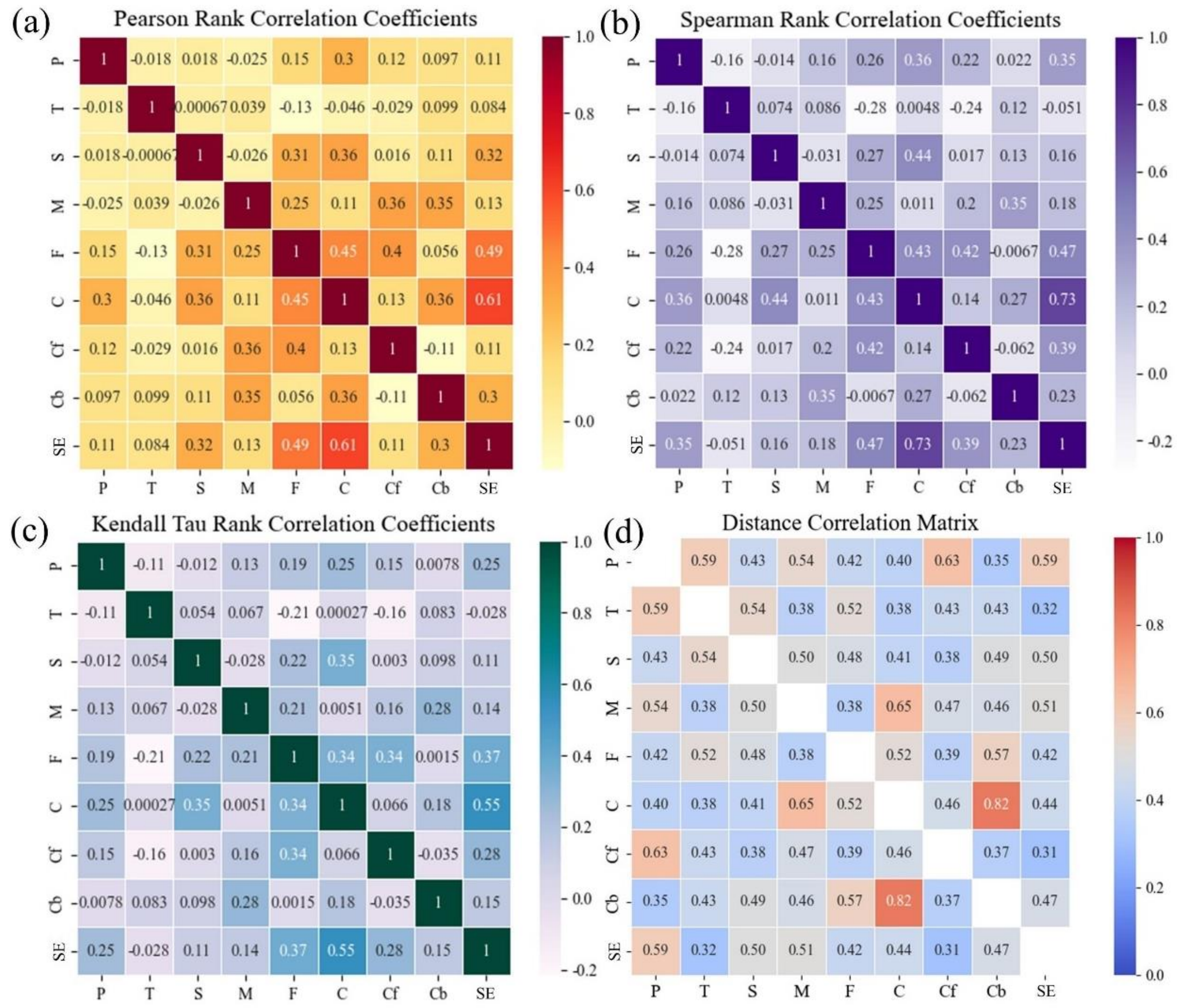

Fig. 4. Feature correlation analysis of electromagnetic interference shielding materials under different measurement methods

(a) Pearson correlation matrix; (b) Spearman correlation matrix; (c) Kendall Tau correlation matrix; (d) Distance correlation matrix

The Pearson correlation matrix, shown in Fig. 4(a), quantifies the linear relationships between features. The results reveal a significant positive correlation between SE and S (correlation coefficient: 0.49), indicating that the sample's structure plays a crucial role in influencing shielding performance. Additionally, there is a strong positive correlation between C and SE, suggesting that higher conductivity contributes to enhanced shielding efficiency. Fig. 4(b) presents the Spearman correlation matrix, where a significant positive correlation is observed between the structural form of F and SE. Furthermore, element types show a positive correlation with structural characteristic, implying that a greater variety of elements promotes more complex structures. As shown in Fig. 4(c), the Kendall Tau rank correlation matrix highlights the ordinal relationships between variables. A notable positive correlation between actual conductivity and SE further supports the reliability of the Pearson analysis. Additionally, a moderate correlation is observed between the proportion of added materials and structural characteristic. Finally, the distance correlation matrix, presented in Fig. 4(d), evaluates both linear and nonlinear relationships between features. It reveals a strong correlation between element types and actual conductivity, as well as a strong correlation between structural characteristic and SE.

By utilizing machine learning data analysis, key electromagnetic shielding performance characteristics in the complex feature space of EMI shielding materials are identified. These characteristics include material conductivity, sample thickness, added elements, and their respective proportions. The screened and collected features, along with the analysis results, align closely with theoretical expectations, indicating that the extracted data features and analysis outcomes are highly accurate. These results can be effectively used to train machine learning models for predicting electromagnetic shielding performance. However, A large number of features may reduce the model's predictive accuracy. Feature optimization is necessary.

In this study, feature optimization is integrated with the laser cladding method. Since the actual conductivity of the sample is unknown prior to laser processing, it is not considered as a feature option. Furthermore, with practical application value and manufacturing costs in mind, the final selected features for laser cladding are the alloy composition (M) and laser energy density (ε). M will be categorized into alloy composition features, while ε will be defined as a laser processing parameter. By comparing the electrical conductivity of various elements in Table 1 and considering the feature optimization results, the Spearman correlation matrix identifies the key alloy elements as iron (Fe), cobalt (Co), silicon (Si), molybdenum (Mo), nickel (Ni), boron (B), and copper (Cu), with Fe and Co serving as the primary matrix elements. The proportions of each element will be determined through machine learning using relevant data from this alloy system, enabling the creation of a specialized database for laser cladding FeCo-based electromagnetic shielding materials.

**2.2. Alloy composition design**

FeCo-based alloys not only hold promise as electromagnetic shielding materials but also exhibit soft magnetic properties [38, 39]. Their high magnetic permeability, low hysteresis loss, and high saturation magnetization enable them to effectively absorb and guide magnetic fields forming closed magnetic loops on their surfaces to block external electromagnetic interference [40-42]. However, due to limited research on the electromagnetic shielding performance of FeCo-based alloys, metals and alloys with similar properties were selected as the basis for component design, laying the groundwork for further investigations into the electromagnetic shielding performance of FeCo-based materials [43-84].

After optimization, Fe, Co, Ni, Mo, Cu, Si, B, and rare earth elements were selected as composition features for the model. Due to the low proportion of rare earth elements in the collected alloy composition data, these elements were considered as optional features to be included in model training when analyzing alloy samples containing rare earth components. Additionally, maximum saturation magnetization ($M_s$, emu/g) and coercivity ($H_c$, Oe) were chosen as output features, while laser energy density (ε, J/mm$^2$) was selected as an input feature, as shown in Table 2.

Table 2 Features of machine learning

| Input Features | | | Output Features |
|---|---|---|---|
| Fe | Mo | B | $M_s$ |
| Co | Cu | （Rare earth） | $H_c$ |
| Ni | Si | ε | — |

The importance ranking of features is shown in Fig. 5. It is evident that the element types and laser energy density significantly influence the $M_s$ and $H_c$ of magnetic materials. The results indicate that Fe and Co have a substantial impact on both $M_s$ and $H_c$, with Fe particularly affecting $M_s$. Boron was found to have the most significant effect on $H_c$, while ε also influences both $M_s$ and $H_c$ to a lesser extent.

Table 3 lists the range of data variations for seven composition features, one process feature, and two property features across 110 data sets. Because the data include pure metals, the atomic percentage ranges for Fe and Co are 0~100 at. %, the atomic percentage ranges of constituent elements are as follows: Ni, 2.32-92.70 at. %; Cu, 0.50-80.00 at. %; Mo, 0.63-8.00 at. %; Si, 0-20.00 at. %; and B, 0.03-30.17 at. %.

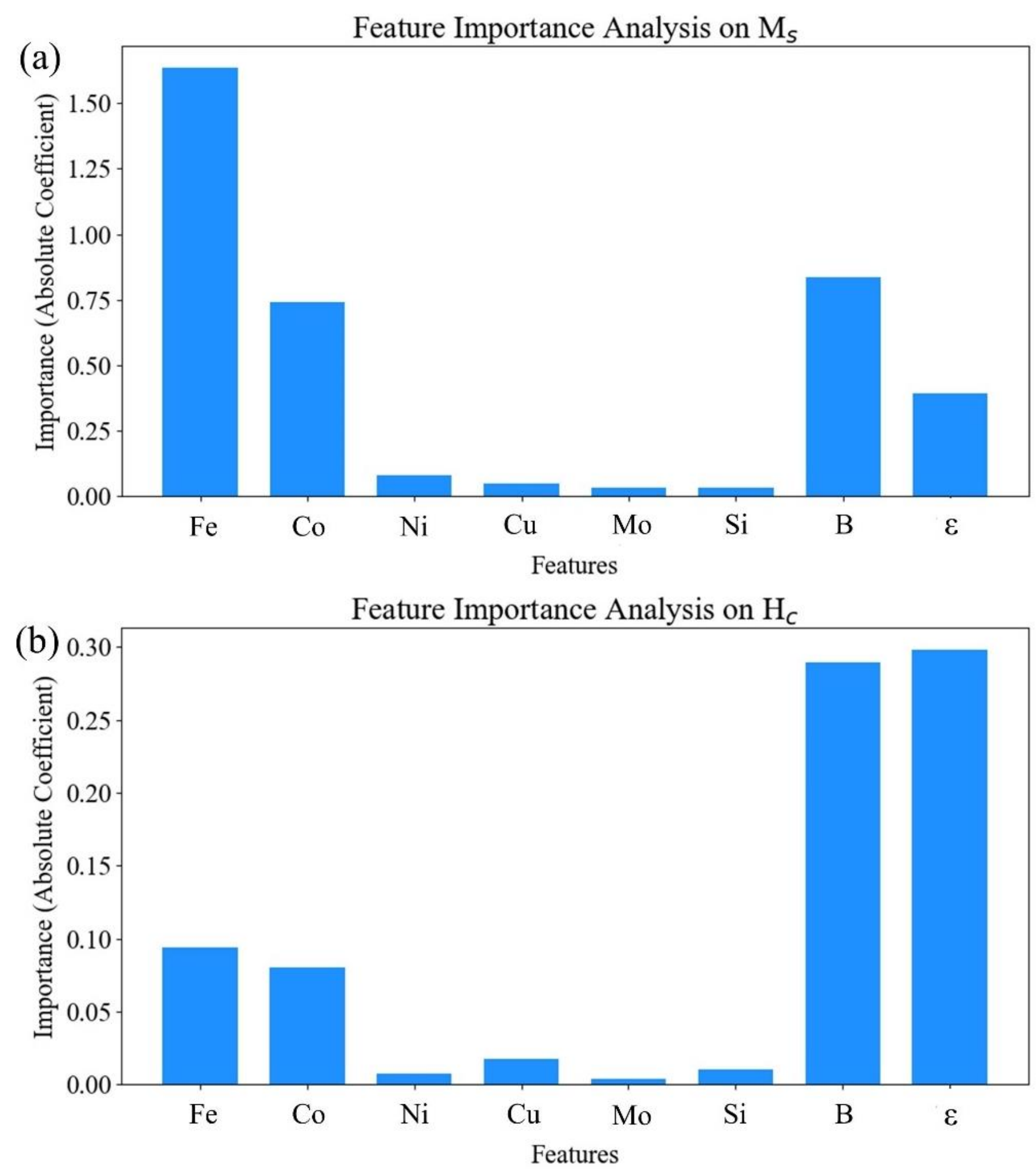

Fig. 5. Comparison of features importance on soft magnetic properties
(a) Importance rank of features influences on saturation magnetization; (b) Importance rank of features influences on coercivity

Table 3 The range of input and output features in the dataset

| Features | Range of variation | |
|---|---|---|
| Fe | 0~100 (at. %) | 0~100 (wt. %) |
| Co | 0~100 (at. %) | 0~100 (wt. %) |
| Ni | 2.32~92.70 (at. %) | 2.00~93.74 (wt. %) |
| Cu | 0.50~80.00 (at. %) | 0.62~81.18 (wt. %) |
| Mo | 0.63~8.00 (at. %) | 1.33~14.43 (wt. %) |
| Si | 0~20.00 (at. %) | 0.12~11.00 (wt. %) |
| B | 0.03~30.17 (at. %) | 0.01~7.68 (wt. %) |
| ε ($J/mm^2$) | 50~90 | |
| $M_s$ (emu/g) | 11.9~248.3 | |
| $H_c$ (Oe) | 4.5~210.0 | |

The frequency density distribution of the elemental features in the magnetic database is shown in Fig. 6. The horizontal axis represents the range of each feature, while the vertical axis represents the frequency density.

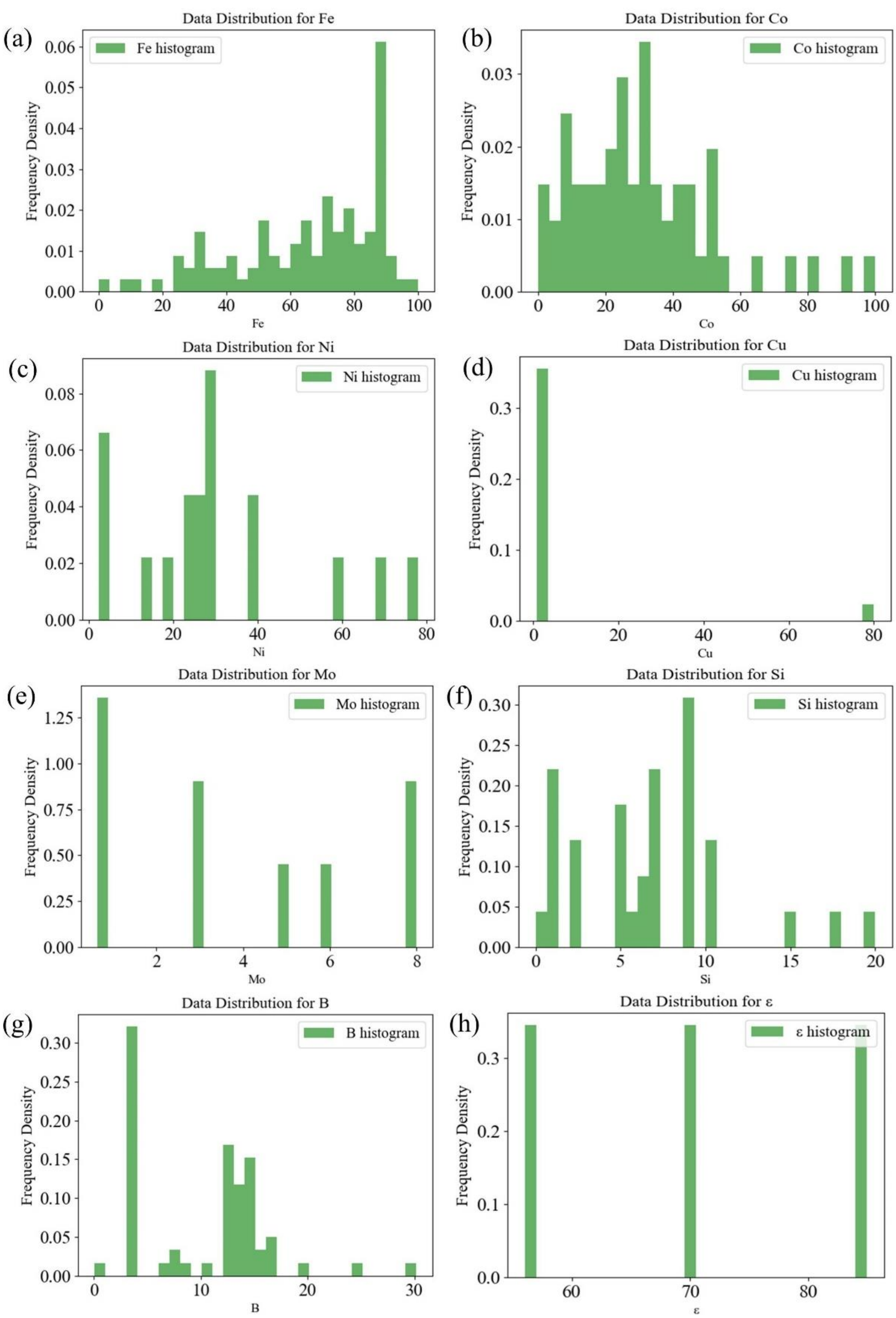

Fig. 6 The distribution of energy density of each element and surface in the dataset

(a) Fe; (b) Co; (c) Ni; (d) Cu; (e) Mo; (f) Si; (g) B; (h) ε

The results indicate that the content distribution of Fe, Co, and B is relatively concentrated. Specifically, the atomic percentage of Fe is primarily between 50~75 at. %, while Co is mostly concentrated between 5~35 at. %. The content distribution of Cu is highly polarized, with fewer data points in the high-content region. Its atomic percentage mainly falls between 0.5~5 at. %, suggesting that Cu is not a primary matrix element. The distributions of Ni and Mo show multiple small peaks, indicating a certain degree of dispersion. Additionally, the content of Si and ε also exhibits dispersion, suggesting potential areas for optimization.

After comparing the importance of the aforementioned features and the distribution ranges of each element, we optimized the compositions and their contents to achieve high saturation magnetization and low coercivity. The specific optimization process is as follows:

(1) Sort the results of $M_s$ for each composition from high to low.

(2) Sort the results of $H_c$ for each composition from low to high.

(3) Within the range where $M_s$ is concentrated, select the compositions with the lowest $H_c$ under high, medium and low saturation magnetization, forming the composition search space as shown in Table 4.

(4) Use the itertools module in Python to iteratively divide the selected composition space. For the Mo element, the minimum and maximum contents are set to 0.5% and 2.0%, respectively, with a step size of 0.1% (i.e., an atomic percentage of 0.1 at. %). This results in 16 possible values: 0.5%, 0.6%, 0.7%, 0.8%, 0.9%, 1.0%, 1.1%, 1.2%, 1.3%, 1.4%, 1.5%, 1.6%, 1.7%, 1.8%, 1.9%, and 2.0%. The values for other elements are determined similarly.

(5) After composition space search, 22 candidate compositions were screened out from 282242 compositions.

After converting atomic percentages to mass percentages, the preliminary optimized design of FeCo-based alloy composition ranges is presented in Table 5. Additionally, research suggests that excessive addition of Si and B elements during laser cladding can degrade microstructural properties [85, 86], while rare earth elements can refine grains and improve formability [87-89]. The selected compositional materials will employ laser cladding technology and rare earth elements for synergistic regulation of microstructural properties. The proportion of Si and B elements will be determined on the content of rare earth elements.

Table 4 Composition search space of FeCo-based alloys

| Element | Min | Max | Step size | Step |
|---|---|---|---|---|
| Fe | 56.0 | 59.0 | 0.1 | 41 |
| Co | 25.0 | 30.0 | 0.1 | 51 |
| Ni | 2.0 | 4.0 | 0.1 | 21 |
| Cu | 0.5 | 2.5 | 0.1 | 21 |
| Mo | 0.5 | 2.0 | 0.1 | 16 |
| Si | 1.0 | 6.0 | 0.1 | 41 |
| B | Bal. | Bal. | — | — |

Table 5 The composition screening results of FeCo-based alloys

| Element | Quality score (wt. %) | Element | Quality score (wt. %) |
|---|---|---|---|
| Fe | 56.0~58.5 | Mo | 1.5~2.0 |
| Co | 27.0~29.5 | Si | 1.5~6.0 |
| Ni | 2.5~4.0 | B | 0.5~2.5 |
| Cu | 1.0~2.0 | Rare earth | Bal. |

The design and manufacturing process of FeCo-based electromagnetic shielding alloys is shown in Fig. 7. An appropriate amount of rare earth elements and their oxides can purify the melt pool, refine grains, and generate functional phases in situ during the laser cladding process which can improve formability, soft magnetic properties, and electromagnetic shielding effectiveness [90-93]. Based on the types of rare earth compounds in the database, $Y_2O_3$, $La_2O_3$, $Nd_2O_3$, and Gd were selected as compositional features to regulate the microstructure and properties of the laser-cladded FeCo-based alloys. Fig. 7 illustrates the optimization process of the FeCo-based electromagnetic shielding alloy composition:

(1) Machine learning is employed to identify critical parameters in electromagnetic shielding materials, including conductivity, thickness, and compositional characteristics. Subsequently, metallic feature extraction is implemented to screen alloy systems compatible with laser cladding processing.

(2) Based on soft magnetic properties, combining the results of feature importance analysis and the characteristics of laser cladding experiments to construct a composition search space. Identify the main elements and their content ranges in FeCo-based alloy system for electromagnetic shielding.

(3) Collect functional phases from the electromagnetic shielding database. Screen these

compounds based on the characteristics of laser cladding, ultimately decide to introduce the above four rare earth elements (Y, La, Nd and Gd) into the FeCo-based alloy system.

(4) Considering the non-equilibrium solidification during laser additive manufacturing and the impact of rare earth elements, construct a new FeCo-based electromagnetic shielding alloy system. The FeCo-based alloy system consists mainly of Fe, Co, Ni, Mo, Cu, Si, B, Y, La, Nd and Gd. The composition of new FeCo-based electromagnetic shielding alloy is shown in Table 6.

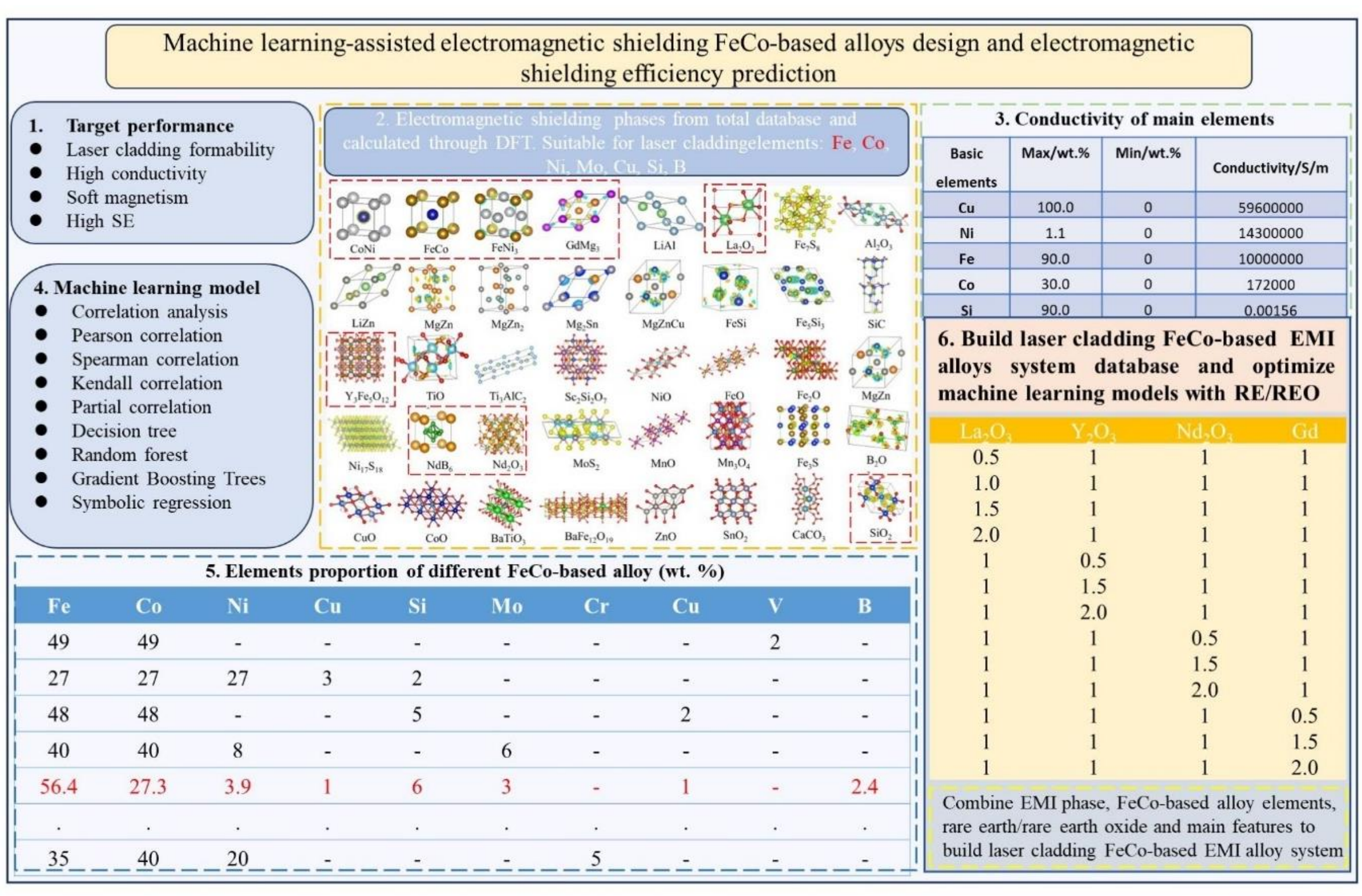

Fig. 7. Machine learning-assisted laser cladding FeCo-based EMI shielding alloys design

Because laser cladding is dynamic non-equilibrium process, samples are prone to generate porosity and cracking during laser cladding, which is detrimental to their soft magnetic properties and electromagnetic shielding performance. Therefore, it is necessary to optimize composition and process parameters. The laser cladding parameters mainly include laser power and scanning speed. By adjusting these parameters, the energy density can be controlled. Energy density is considered a critical process characteristic and plays a key role in enhancing the experimental database for laser cladding new FeCo-based electromagnetic shielding alloys.

Table 6 Composition of FeCo-based alloys compounded with rare earth and rare earth oxide（wt. %）

| Sample name | Fe | Co | Ni | Cu | Mo | Si | B | $Y_2O_3$ | $La_2O_3$ | $Nd_2O_3$ | Gd |
|---|---|---|---|---|---|---|---|---|---|---|---|
| FeCo-0$Y_2O_3$ | 56.4 | 27.3 | 3.9 | 1.0 | 3.0 | 6.0 | 2.4 | 0 | 0 | 0 | 0 |
| FeCo-0.5$Y_2O_3$ | 56.4 | 27.3 | 3.9 | 1.0 | 3.0 | 5.5 | 2.4 | 0.5 | 0 | 0 | 0 |
| FeCo-1.0$Y_2O_3$ | 56.4 | 27.3 | 3.9 | 1.0 | 3.0 | 5 | 2.4 | 1.0 | 0 | 0 | 0 |
| FeCo-1.5$Y_2O_3$ | 56.4 | 27.3 | 3.9 | 1.0 | 3.0 | 4.5 | 2.4 | 1.5 | 0 | 0 | 0 |
| FeCo-2.0$Y_2O_3$ | 56.4 | 27.3 | 3.9 | 1.0 | 3.0 | 4.0 | 2.4 | 2.0 | 0 | 0 | 0 |
| FeCo-0Y1La1Nd1Gd | 58.5 | 29.3 | 2.9 | 2.0 | 2.0 | 2.0 | 0.5 | 0 | 1.0 | 1.0 | 1.0 |
| FeCo-0.5Y1La1Nd1Gd | 58.3 | 29.1 | 2.9 | 1.9 | 1.9 | 1.9 | 0.5 | 0.5 | 1.0 | 1.0 | 1.0 |
| FeCo-1Y1La1Nd1Gd | 58.0 | 29.0 | 2.9 | 1.9 | 1.9 | 1.9 | 0.5 | 1.0 | 1.0 | 1.0 | 1.0 |
| FeCo-1.5Y1La1Nd1Gd | 57.7 | 28.8 | 2.9 | 1.9 | 1.9 | 1.9 | 0.5 | 1.4 | 1.0 | 1.0 | 1.0 |
| FeCo-2Y1La1Nd1Gd | 57.4 | 28.7 | 2.9 | 1.9 | 1.9 | 1.9 | 0.5 | 1.9 | 1.0 | 1.0 | 1.0 |
| FeCo-1Y0La1Nd1Gd | 58.5 | 29.3 | 2.9 | 2.0 | 2.0 | 2.0 | 0.5 | 1.0 | 0 | 1.0 | 1.0 |
| FeCo-1Y0.5La1Nd1Gd | 58.3 | 29.1 | 2.9 | 1.9 | 1.9 | 1.9 | 0.5 | 1.0 | 0.5 | 1.0 | 1.0 |
| FeCo-1Y1.5La1Nd1Gd | 57.7 | 28.8 | 2.9 | 1.9 | 1.9 | 1.9 | 0.5 | 1.0 | 1.4 | 1.0 | 1.0 |
| FeCo-1Y2La1Nd1Gd | 57.4 | 28.7 | 2.9 | 1.9 | 1.9 | 1.9 | 0.5 | 1.0 | 1.9 | 1.0 | 1.0 |
| FeCo-1Y1La0Nd1Gd | 58.5 | 29.3 | 2.9 | 2.0 | 2.0 | 2.0 | 0.5 | 1.0 | 1.0 | 0 | 1.0 |
| FeCo-1Y1La0.5Nd1Gd | 58.3 | 29.1 | 2.9 | 1.9 | 1.9 | 1.9 | 0.5 | 1.0 | 1.0 | 0.5 | 1.0 |
| FeCo-1Y1La1.5Nd1Gd | 57.7 | 28.8 | 2.9 | 1.9 | 1.9 | 1.9 | 0.5 | 1.0 | 1.0 | 1.4 | 1.0 |
| FeCo-1Y1La2Nd1Gd | 57.4 | 28.7 | 2.9 | 1.9 | 1.9 | 1.9 | 0.5 | 1.0 | 1.0 | 1.9 | 1.0 |
| FeCo-1Y1La1Nd0Gd | 58.5 | 29.3 | 2.9 | 2.0 | 2.0 | 2.0 | 0.5 | 1.0 | 1.0 | 1.0 | 0 |
| FeCo-1Y1La1Nd0.5Gd | 58.3 | 29.1 | 2.9 | 1.9 | 1.9 | 1.9 | 0.5 | 1.0 | 1.0 | 1.0 | 0.5 |
| FeCo-1Y1La1Nd1.5Gd | 57.7 | 28.8 | 2.9 | 1.9 | 1.9 | 1.9 | 0.5 | 1.0 | 1.0 | 1.0 | 1.4 |
| FeCo-1Y1La1Nd2Gd | 57.4 | 28.7 | 2.9 | 1.9 | 1.9 | 1.9 | 0.5 | 1.0 | 1.0 | 1.0 | 1.9 |

## 3. Laser cladding process design

To adjust the laser power and scanning speed for FeCo-based alloys with different elemental compositions, the ranges are set at 2000~2900 W for laser power and 3~10 mm/s for scanning speed (laser energy density range: 56~140 J/mm$^2$). The goal is to minimize defects and improve EMI shielding efficiency. The laser cladding process parameters are primarily determined based on the Dlight-3000 W semiconductor laser system and material properties. By combining the new FeCo-based electromagnetic shielding alloy system with these process parameters, a database consisting of 147 sets (Supplementary Table S1) has been constructed.

## 4. Machine learning model establishment and evaluation

### 4.1. Establishment of a database for laser cladding FeCo-based alloys

The quality of input features plays a decisive role in the performance of machine learning models. Based on previous studies optimizing the microstructure and properties of laser-cladded FeCo-based alloys, 147 sets of experimental data were collected. The details of these features are listed in Table 7. The main preprocessing methods include unit normalization, data cleaning, and feature dimensionality reduction. In this work, the input features consist of elemental proportions and laser energy density, providing simple and efficient references for predicting target performance. The output features primarily include electromagnetic shielding performance, with coercivity and saturation magnetization as alternative features.

Table 7 Features of laser cladding FeCo-based alloys

| Input Features | | | Output Features |
|---|---|---|---|
| Fe | Cu | La（$La_2O_3$） | SE |
| Co | Si | Nd（$Nd_2O_3$） | $M_s$ |
| Ni | B | Gd | $H_c$ |
| Mo | Y（$Y_2O_3$） | ε | — |

**4.2. Feature engineering and dimensionality reduction**

Data analysis revealed that some features in the FeCo-based alloy systems have less relevance. Increasing the feature dimensions leads to a larger data space, making it more challenging for the model to capture useful information effectively. Therefore, dimensionality reduction is necessary to simplify the model, reduce training time, and enhance accuracy. First, a correlation analysis is conducted on the initial features of the FeCo-based alloys. Then, features are optimized based on the correlation results. Since there is a non-linear relationship between composition, laser parameter, and EMI shielding performance, a distance correlation matrix is used for analysis.

Fig. 8 presents the results of the distance correlation matrix for the FeCo-based alloy systems. The color coding in the correlation matrix indicates the strength of the correlations between features: dark red represents a strong positive correlation, dark blue indicates a strong negative correlation, and gray signifies no significant or weak correlation. By analyzing the distance correlation matrix, the relationships between $M_s$, $H_c$, and SE with other features can be identified. $M_s$ correlates with Fe, Co and Cu, suggesting that as the content of these elements increases, $M_s$ also tends to rise significantly. $H_c$ shows a notable positive correlation with Co, Cu, Gd and ε (correlation coefficients: 0.47, 0.51, 0.45 and 0.50, respectively), indicating that increases in these features lead to higher $H_c$. SE exhibits a moderate positive correlation with Fe, Cu, B and Gd (correlation coefficients: 0.49, 0.39, 0.44 and 0.30, respectively). Considering the model's dimensionality, the correlations between features, and the variation range in elemental features, B and Ni are excluded to ensure effective dimensionality while maintaining model accuracy.

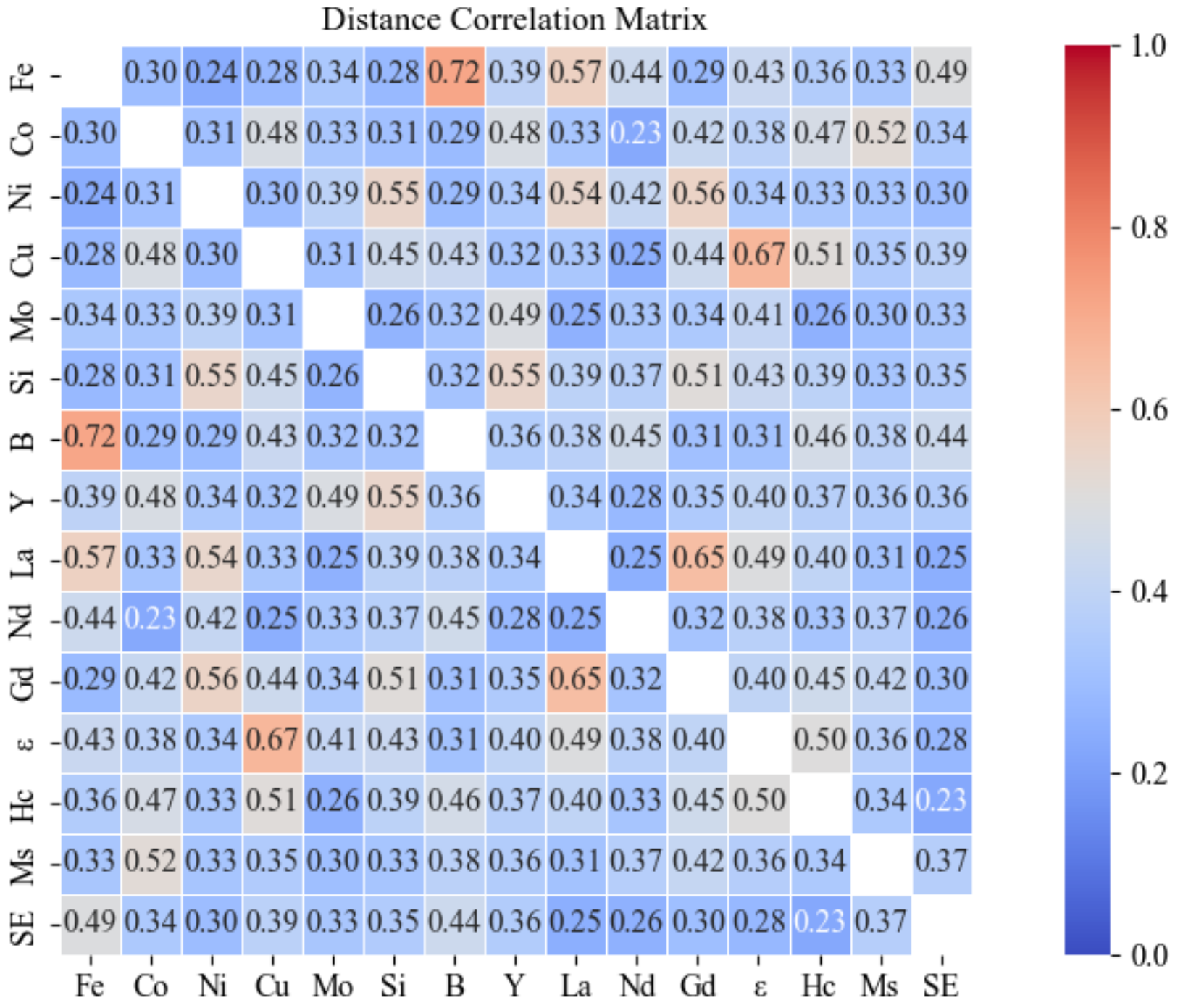

Fig. 8. Initial feature correlation analysis of laser-cladded FeCo-based alloys

Fig. 9 illustrates the thermodynamic distribution of the distance correlation coefficients after dimensionality reduction. As shown in Fig. 9, the distance correlation matrix can capture both linear and nonlinear relationships, particularly when complex dependencies are present. The results indicate that the relationships between features are intricate and cannot be directly described. This mirrors the non-equilibrium nature of laser cladding, where the complex interactions between alloy compositions and laser processing introduce various uncertainties. Directly obtaining all structural information from the non-equilibrium reaction process is not feasible. Machine learning methods help simplify the intermediate stages of laser cladding, enabling direct prediction of target performance from initial parameters. This approach can significantly reduce the experimental cycle for developing new FeCo-based electromagnetic shielding alloys.

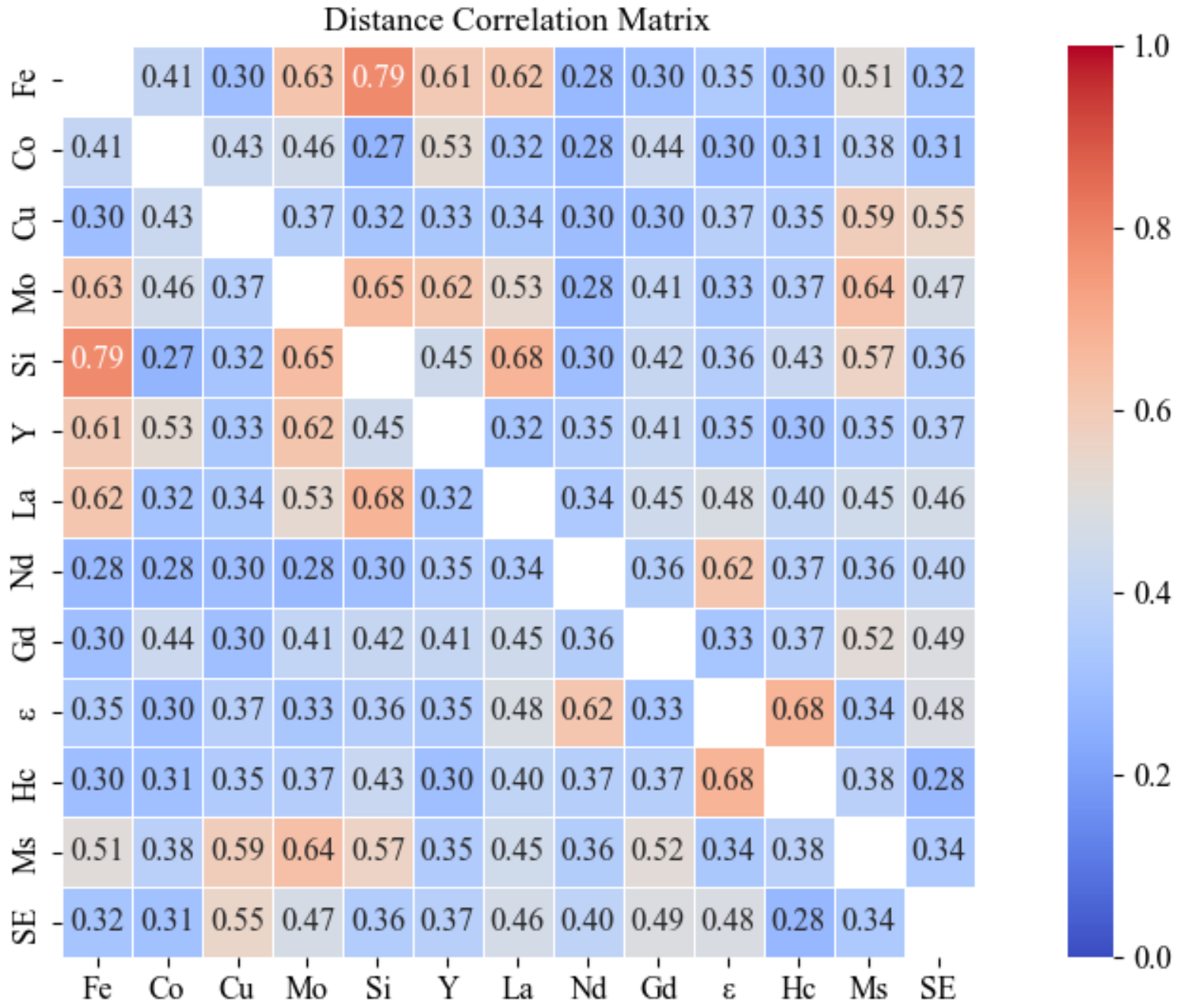

Fig. 9. Feature dimension reduction of distance correlation coefficients

### 4.3. Machine learning model evaluation

After selecting the key features from laser-cladded FeCo-based electromagnetic shielding alloys, different machine learning models were employed to predict the EMI shielding performance. The prediction accuracies of these models were analyzed and compared. Six algorithms were integrated: Decision Tree, Random Forest, Support Vector Regression (SVR), Ridge Regression, K-Nearest Neighbors (K-NN), and Gradient Boosting [94-102]. To enhance performance, the Gradient Boosting algorithm was further improved with Extreme Gradient Boosting (XGBoost), Light Gradient Boosting Machine (LightGBM) and CatBoost. The machine learning methods and corresponding initial parameters are outlined in Table 8.

Table 8 Nine machine learning algorithms and initial parameter settings

| Algorithm | Initial parameters |
| --- | --- |
| Decision Tree | 'max_depth': [10, 20, 30] |
| Random forest | 'max_depth': [10, 20, 30] |
| SVR | 'C': [1, 10, 100] |
| Ridge Regression | 'alpha': [1, 10, 100, 1000] |
| K-Nearest Neighbors | 'n_neighbors': [3, 5, 7, 10] |
| Gradient Boosting | 'max_depth': [3, 5, 7] |
| XGBoost | 'max_depth': [3, 6, 9] |
| LightGBM | 'num_leaves': [31, 63, 127] |
| Catboost | 'depth': [4, 6, 8] |

The evaluation metrics for prediction accuracy primarily include Mean Absolute Error (MAE), Root Mean Square Error (RMSE), and Coefficient of Determination ($R^2$) [103-105]. MAE represents the average absolute error between the predicted and actual values, providing a direct reflection of the average prediction error. RMSE, the square root of the mean of squared differences between predicted and actual values, tends to amplify larger deviations in the model. $R^2$ is a statistical measure used to assess model performance, indicating the proportion of variance in the dependent variable that can be explained by the independent variables. Its value ranges from 0 to 1. The database for the laser-cladded FeCo-based electromagnetic shielding alloys was divided into training set and test set. Table 9 presents the test set results for predicting EMI shielding effectiveness through different machine learning models.

Table 9 Accuracy comparison of different machine learning methods for predicting SE

| Model | Model accuracy evaluation index | | |
|---|---|---|---|
| | $R^2$ | MAE | RMSE |
| Decision Tree | 0.72 | 8.54 | 14.74 |
| Random forest | 0.82 | 7.27 | 11.66 |
| SVR | 0.83 | 7.61 | 11.32 |
| Ridge Regression | 0.75 | 10.3 | 13.87 |
| K-NN | 0.84 | 7.12 | 10.93 |
| Gradient Boosting | 0.82 | 7.71 | 11.82 |
| XGBoost | 0.83 | 7.56 | 11.39 |
| LightGBM | 0.84 | 7.55 | 11.24 |
| Catboost | 0.81 | 7.87 | 12.00 |

Additionally, K-Fold Cross-Validation and GridSearch were applied to optimize and evaluate the prediction performance of the models. The dataset was partitioned into five mutually exclusive subsets using the 5-fold cross-validation method. In each iteration, one subset was designated as the validation set, while the remaining four subsets were used as the training set, allowing for a comprehensive evaluation of the model's generalization performance across all data partitions. This approach effectively reduces the risk of overfitting. Furthermore, GridSearch was utilized to perform a grid search for the hyperparameters of each model. Fig. 10 illustrates the test set performance of nine machine learning models for predicting SE.

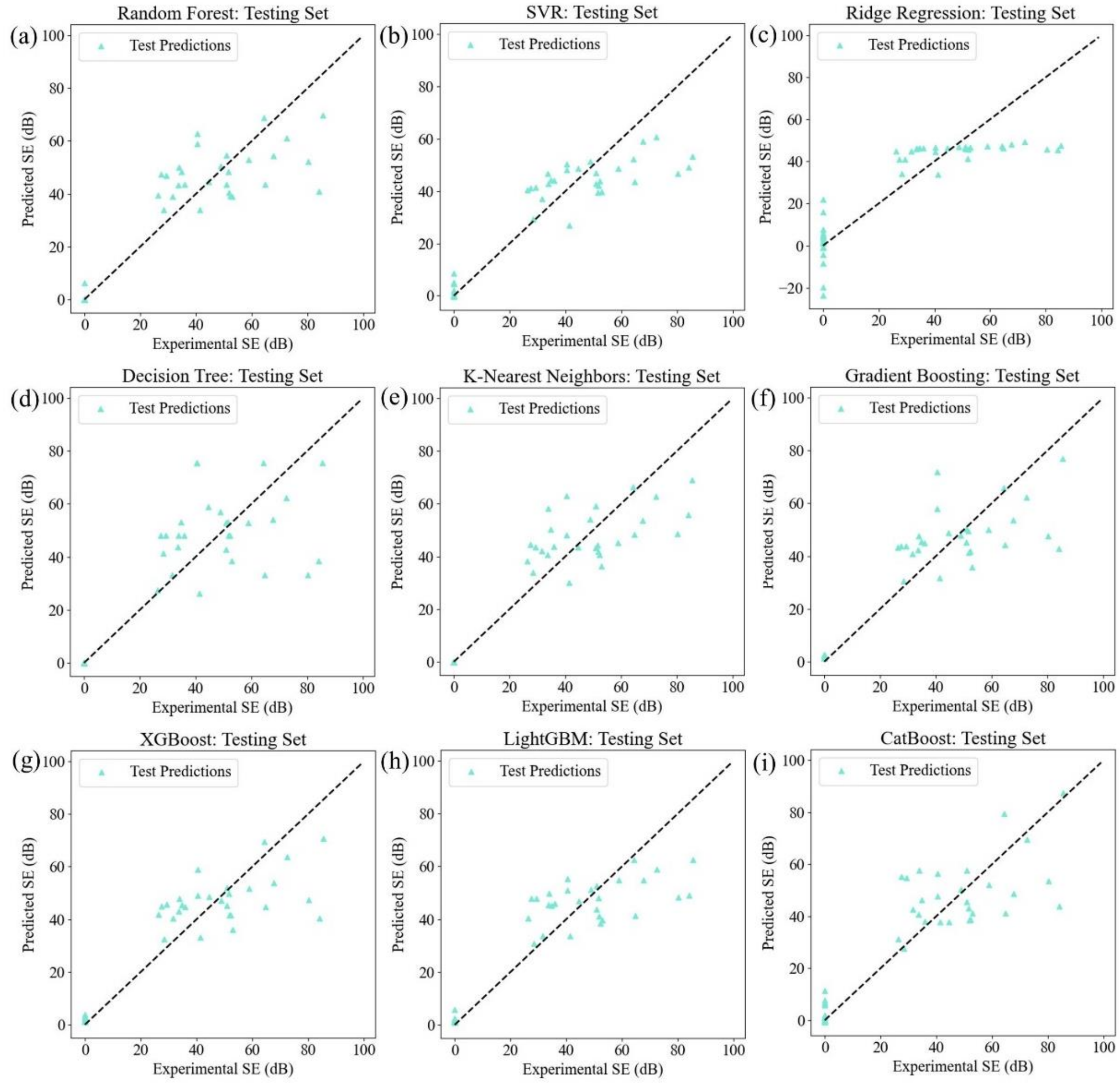

Fig. 10. Comparison of nine machine learning models in predicting electromagnetic SE

(a) Random Forest; (b) SVR; (c) Ridge Regression; (d) Decision Tree; (e) K-NN; (f) Gradient Boosting; (g) XGBoost; (h) LightGBM; (i) CatBoost

Based on the evaluation results, the K-NN algorithm demonstrated the best predictive performance, achieving the highest $R^2$ (0.84), and the lowest RMSE (10.93) and MAE (7.12). In contrast, the Decision Tree and Ridge Regression models performed less effectively, with larger prediction errors and weaker model interpretability. Table 10 compares the experimental values with the K-NN model's predicted values for electromagnetic shielding effectiveness of the laser-cladded FeCo-based alloys in the test set. The input features include different element contents and laser energy density, while SE serves as the output feature. The $R^2$ value of the K-NN model on the test set reached 0.84.

Overall, after dimensionality reduction and feature optimization for the laser-cladded FeCo-based alloys, the K-NN model effectively captures the feature information in the database, demonstrating strong prediction accuracy and generalization ability. The discrepancies between the predicted and experimental results suggest that the current computational model needs further refinement, thus improving its predictive accuracy and practical applicability.

Table 10 Experimental SE vs. K-NN model predicted SE of FeCo-based alloys in the test set

| Sample name | Experimental SE | Predicted SE |
|---|---|---|
| FeCo-2Y1La1Nd1Gd-73 | 28 | 34 |
| FeCo-0Y1La1Nd1Gd-68 | 52 | 42 |
| FeCo-1Y1.5La1Nd1Gd-68 | 32 | 42 |
| FeCo-1Y1La1Nd0.5Gd-68 | 36 | 44 |
| FeCo-1Y1La1Nd0Gd-81 | 64 | 66 |
| FeCo-1Y1La1Nd2Gd-73 | 84 | 56 |
| FeCo-1Y0La1Nd1Gd-82 | 72 | 63 |
| FeCo-1Y1La1Nd1Gd-84 | 51 | 59 |
| FeCo-1Y1La0.5Nd1Gd-66 | 52 | 44 |
| FeCo-1Y1La1Nd1Gd-69 | 51 | 43 |
| FeCo-1Y0.5La1Nd1Gd-68 | 49 | 54 |
| FeCo-1Y1.5La1Nd1Gd-84 | 26 | 38 |
| FeCo-1Y1La1Nd1.5Gd-78 | 34 | 41 |
| FeCo-1Y1La1.5Nd1Gd-69 | 52 | 41 |
| FeCo-1Y1La1Nd1.5Gd-84 | 45 | 44 |
| FeCo-1Y1La1Nd0Gd-84 | 85 | 69 |
| FeCo-1Y1La1Nd0Gd-104 | 40 | 48 |
| FeCo-1Y0La1Nd1Gd-72 | 68 | 54 |

### 4.4. Φ-SO analysis for electromagnetic shielding performance

To extend the applicability of machine learning algorithms to various types of electromagnetic shielding materials, general features were extracted from laser-cladded FeCo-based alloys and other types of electromagnetic shielding materials. The input features for the Φ-SO model included the actual conductivity of the samples, the theoretical conductivity of the added components, the theoretical conductivity of the matrix, and the structural grade of the added components (classified as powder, hollow tubular, solid wire, pure solid, and polymorphic mix). Among these, the rare earth component is in powder form. Due to the significant numerical range differences among the features, data normalization and standardization were performed before training the Φ-SO model. The mathematical expression derived from the Φ-SO model is applicable to FeCo-based alloys, achieving an $R^2$ value of 0.80. The theoretical relationship between SE and the features is expressed by the following formula:

$$SE = (C + F) \times C \times Cf\sqrt{Cb \times F}$$

In the formula, *C* represents the actual conductivity of the laser-cladded FeCo-based alloy (S/m); *F* denotes the structural grade of rare earth elements; *Cf* is the theoretical conductivity of the rare earth elements (S/m); and *Cb* is the theoretical conductivity of the FeCo-based alloy (S/m).

Φ-SO offers several significant advantages. First, it utilizes global optimization techniques, ensuring the global optimality of the search process. Second, Φ-SO generates more concise and meaningful expressions, enhancing the interpretability of the model. Additionally, Φ-SO is highly adaptable, capable of adjusting to different data types and functions. The relationship formula between these physical features and electromagnetic shielding performance is applicable to other electromagnetic shielding materials. However, the accuracy of the model still requires further improvement.

## 5. Conclusions

A new multifunctional alloy system for laser cladding has been established by integrating a soft magnetic alloy database with EMI characteristics and laser additive manufacturing technology. Through the synergistic use of machine learning algorithms, high-dimensional compositional space exploration, and laser cladding process parameterization, 22 rare earth-modified FeCoNiCuMoSiB alloy compositions were optimized from a pool of 282242 candidate systems. Additionally, a specialized database containing 147 parameter sets was created for new FeCo-based alloys with rare earth additions, achieved by correlating energy density metrics (18~35 $J/mm^2$) with elemental distribution characteristics through a systematic laser cladding process design.

Multiple machine learning models (Decision Tree, Random Forest, SVR, Ridge Regression, K-NN, and Gradient Boosting) were employed to predict EMI shielding efficiency based on initial elements, conductivity, and laser cladding parameters. The laser cladding parameters were characterized by energy density, and the explainability of the machine learning models was further enhanced using Φ-SO. The predicted results were compared with experimental results, and the performance of different models was evaluated using $R^2$, MAE, and RMSE. The results indicate that the K-NN model achieved superior prediction accuracy ($R^2$: 0.84, RMSE: 10.93, MAE: 7.12) within the same dataset. Additionally, Φ-SO improved the interpretability of the machine learning models by deriving expressions that describe the relationships between various physical quantities, thus aiding in understanding the mathematical correlations between different features. The predictive accuracy of the model holds potential for further improvement through continued expansion and refinement of the materials database.

## 6. Materials and methods

### 6.1. Materials representation

The average diameter of the elemental powders was approximately 74 μm. Powder blending was performed using a QM-5 planetary ball mill, with process parameters optimized to a rotational speed of 300 rpm for 8 hours, using stainless steel grinding media and a ball-to-powder ratio of 5:1. Laser deposition was carried out with an FL-DLS21 fiber laser system (FLLC-3000), producing a rectangular focal spot of 4×4 $mm^2$.

### 6.2. Machine learning models

The following machine learning models were applied: Decision Tree, Random Forest, Support Vector Regression (SVR), Ridge Regression, K-Nearest Neighbors (K-NN), and Gradient Boosting. The Gradient Boosting algorithm was further enhanced to include Extreme Gradient Boosting (XGBoost), Light Gradient Boosting Machine (LightGBM), and CatBoost. The Φ-SO symbolic regression framework integrates a hybrid deep reinforcement learning architecture to uncover potential physical laws from sparse datasets, performing combinatorial optimization within multi-dimensional operator spaces.

**Acknowledgments Funding**: This work was financially supported by National Key R&D Program of China(2016YFB1100201), Green Manufacturing System Integration Project of the Industry and Information Ministry of China (2017), Shenyang important scientific and technological achievements transformation project (20-203-5-6). **Author contributions:** Special thanks are due to the instrumental or data analysis from Analytical and Testing Center (Northeastern University). **Competing interests**: The authors declare that they have no competing interests.

**Data and materials availability**: All data needed to evaluate the conclusions in the paper are present in the paper and/or the Supplementary Materials. Data are available in a GitHub reposi tory (https://github.com/liuzyzju/phonon_TL). Additional data related to this paper may be req uested from the authors.